\documentclass{ws-ijmpa}

\begin{document}

\markboth{F. A. Chishtie, V. Elias, T. G. Steele}
{Startling Equivalences in the Higgs-Goldstone Sector}

%
\catchline{}{}{}{}{}
%

\title{STARTLING EQUIVALENCES IN THE HIGGS-GOLDSTONE SECTOR BETWEEN RADIATIVE AND LOWEST-ORDER CONVENTIONAL ELECTROWEAK SYMMETRY BREAKING}

\author{F. A. CHISHTIE}

\address{Department of Applied Mathematics,The University of Western Ontario,\\ 1151 Richmond Street North,
London, Ontario N6A 5B7,
Canada\\
fachisht@uwo.ca}

\author{V. ELIAS}

\address{Department of Applied Mathematics, The University of Western Ontario,\\1151 Richmond Street North, London, Ontario N6A 5B7, Canada
\\
velias@uwo.ca}

\author{T. G. STEELE}

\address{Department of Physics and Engineering Physics, University of Saskatchewan,\\Saskatoon, Saskatchewan S7N 5E2, Canada
\\
Tom.Steele@usask.ca}

\maketitle


\begin{abstract}
For the Higgs boson mass of $\sim 220$ GeV expected to arise from radiative electroweak symmetry breaking, we find the same lowest-order expressions as would be obtained from conventional electroweak symmetry breaking, given the same Higgs boson mass, for Higgs-Goldstone sector scattering processes identified with $W_L^ + W_L^- \rightarrow W_L^+ W_L^-$, $W_L^+ W_L^- \rightarrow Z_L Z_L$, as well as for Higgs boson decay widths $H \rightarrow W_L^+ W_L^-$, $H \rightarrow Z_L Z_L$.  The radiatively broken case, however, leads to an order of magnitude enhancement over lowest-order conventional symmetry breaking for scattering processes $W_L^+ W_L^+ \rightarrow H H$, $Z_L Z_L \rightarrow H H$, as well as a factor of $\sim 30$ enhancement for $H H \rightarrow H H$.

\keywords{Higgs boson phenomenology; radiative electroweak  symmetry breaking.}
\end{abstract}

\ccode{PACS numbers: 11.30.Qc, 11.15.Ex, 14.80Bn.} \ccode{Report
No.: UWO-TH-05/01}

\bigskip

In the conventional single-Higgs doublet Standard Model Lagrangian, a scalar field mass term within a unification context would necessarily be subject to GUT-scale radiative corrections which would require fine-tuning for the Higgs mass to remain within its present indirect-measurement bounds. One of the ways to address this problem is to assume that the electroweak Lagrangian is protected from acquiring such a scalar field mass term by some higher symmetry.  The classic Coleman and Weinberg calculation \cite{1} demonstrated that radiative corrections to such a protected Lagrangian would still lead to a spontaneous symmetry breaking in which the Higgs boson mass can be explicitly predicted.  However, this calculation preceded the discovery of the top-quark, whose large Yukawa coupling $g_t$ leads to fermion loop contributions which destabilize the effective potential, unless the scalar field self-interaction coupling $\lambda$ is too large for predictive physics to emerge.  Indeed, it has been argued that the expansion parameter for the large-$\lambda$ case of the Coleman-Weinberg potential is $\lambda \log [(\phi/<\phi>)^2]$, rather than $\lambda$ itself,  \cite{2} in which case the contribution of leading logarithms needs to be controlled to all orders.

Recently it has been demonstrated that if one uses renormalization-group methods to include all such leading logarithms in the effective potential series devolving from the dominant Standard Model coupling constants [$\lambda$, $g_t$, and the $SU(3) \times SU(2) \times U(1)$ gauge coupling constants], one then obtains a Higgs boson mass of $218$ GeV within indirect experimental bounds accompanied by a value for $\lambda$ sufficiently small for the apparent convergence of Standard Model $\beta$- and $\gamma$-functions. \cite{3}

Of course, the Higgs boson masss is an arbitrary parameter in conventional electroweak symmetry breaking (CSB).  Thus, if a $220$ GeV Higgs boson {\it were} found, one could not automatically assume that radiative electroweak symmetry breaking (RSB) had occurred.  To demonstrate radiative electroweak symmetry breaking, one needs to examine collateral phenomenological signatures of radiative symmetry breaking in the Higgs-Goldstone sector.  The enhanced value of $\lambda$ characterizing the radiatively broken case should percolate into enhanced couplings of the Higgs boson to both itself and to the longitudinal components of the weak-interaction gauge bosons $W^\pm$ and $Z$.  Such couplings are examined below to search for realistic signatures of radiative symmetry breaking.
We find, however, that a number of these couplings are found to be the {\it same as} in lowest-order CSB, given the same Higgs boson mass as arises in RSB.

Initially RSB differs from CSB insofar as only the CSB tree potential contains an explicit negative mass term.  The CSB tree potential is
\begin{equation}
V_{CSB} = -\mu^2 \phi^\dagger \phi + \lambda (\phi^\dagger \phi)^2
\label{``(1)''}
\end{equation}
where $\phi$ denotes the complex scalar field doublet
\begin{equation}
\phi = \left[ (\phi_1 + i \phi_2)/\sqrt{2}, \; \; (\phi_3 + i \phi_4)/\sqrt{2}\right] \; .
\label{``(2)''}
\end{equation}
The charged Goldstone fields $(\phi_1 \pm i \phi_2)/\sqrt{2}$ become identified with the longitudinal ($L$) components of the charged electroweak gauge boson fields $W_L^\pm$, and the neutral Goldstone field $\phi_4$ correspondingly disappears in unitary gauge into the logitudinal component of the neutral massive electroweak gauge boson $Z_L$.

$V_{CSB}$ acquires a vacuum expectation value ($vev$)
\begin{equation}
v \equiv \sqrt{\mu^2 / \lambda} = 2M_w/g_2 = 246 \; GeV \; .
\label{``(3)''}
\end{equation}
The physical Higgs boson field is then $H = \phi_3 - v$.  Since the Higgs boson mass is just
\begin{equation}
M_H^2 \equiv V_{CSB}^{\prime\prime}\left. \right|_v \equiv \partial^2 V_{CSB} / \partial \left. H^2\right|_{H = \phi_{1,2,4}=0} = 2\mu^2
\label{``(4)''}
\end{equation}
we see from the definition (3) of $v$ that
\begin{equation}
\lambda_{CSB} = M_H^2 / 2v^2 \; .
\label{``(5)''}
\end{equation}
Thus if CSB resulted in a Higgs boson with $M_H = 218$ GeV, then $\lambda_{CSB} = 0.393$.

This is to be contrasted with RSB, which dictates to leading-logarithm order a Higgs boson mass $M_H = 218$ GeV accompanied by a quartic scalar-field interaction constant  referenced to the $vev$ energy scale $\lambda_{RSB} = 2.15$ ($y(v) \equiv \lambda_{RSB} / 4\pi^2 = 0.0545$ \cite{3}).  Thus RSB, which differs from CSB at the outset only by the absence of the quadratic mass term in eq. (1), is seen to result in a more than five-fold enhancement of $\lambda$ over the CSB case for an equivalent value of $M_H$.

Now at high energies ($s >> M_W^2$), it is known \cite{4} up to terms of order $M_W / \sqrt{s}$ that the scattering matrix for $W_L^\pm$, $Z_L$, and $H$ is the same as that for self-interactions of $( \phi_1 \pm i \phi_2) / \sqrt{2}$, $\phi_4$ and $H$, respectively.  This result is cited elsewhere \cite{5} as the ``Goldstone boson equivalence theorem''. Since the factor quartic in fields is common to the tree potentials for CSB {\it and} RSB, and since at high energies ($s >> M_H^2 / 2$) the term quadratic in fields in the CSB tree-potential (1) is irrelevant, \cite{4} one would necessarily conclude that scattering processes such as $W_L^+ W_L^- \rightarrow (Z_L Z_L, \; W_L^+ W_L^-)$ or the decay rates $H \rightarrow (W_L^+ W_L^- , \; Z_L Z_L)$ would be enhanced in RSB by a factor of $\sim 30$ relative to the processes in CSB, given a fortuitously equivalent value for $M_H$.

Now the Higgs boson equivalence theorem is of particular use when $\lambda$ is large, since corrections higher order in $\lambda$ in a CSB context can be addressed perturbatively in $\lambda$ with neglectable ${\cal{O}}(g_2^2)$ corrections.
A renormalization scheme has been developed \cite{5} for two-loop CSB calculations in which the constant of proportionality between scattering amplitudes $T(W_L^\pm, \; Z_L)$ and corresponding Goldstone-boson sector amplitudes $T\left( (\phi_1 \pm i \phi_2) / \sqrt{2}, \; \phi_4 \right)$ is $\left[ 1 + {\cal{O}} (g^2, g g^\prime, g^{\prime^2}) \right]$.  This approach is further refined for the CSB case by Riesselmann and collaborators \cite{5,6} to obtain high energy expressions for such amplitudes when $M_H$, and concomitantly $\lambda_{CSB}$, is large.

For the case of RSB, all such Goldstone-boson sector amplitudes can be extracted from the effective potential generated by Yukawa, scalar and gauge interactions \cite{1,3}.
\begin{equation}
V_{RSB} = 4\pi^2 (\phi^+ \phi)^2 \left[ A + BL + CL^2 + DL^3 + EL^4 + \ldots \right]
\label{``(6)''}
\end{equation}
where
\begin{equation}
L \equiv \log \left[ 2 \phi^+ \phi / v^2 \right] = \log \left[ \sum_{i = 1}^4 \phi_i^2 / v^2\right],
\label{``(7)''}
\end{equation}
and where the constants $\{A-E\}$ in eq. (6) are determined by renormalization-group methods as polynomial functions of the dominant Standard Model coupling-constants $\{\lambda, g_t, g_3, g_2, g^\prime \}$.  Using the notation $|_v$ to mean $\phi_3 = v, \phi_1 = \phi_2 = \phi_4 = 0$, the minimization requirement \cite{1}
\begin{equation}
\left. V_{RSB}^\prime \right|_v = 0
\label{``(8)''}
\end{equation}
implies that $A = -B/2$.  The Higgs boson mass is found from the second derivative condition
\begin{equation}
\left. \frac{\partial^2 V_{RSB}}{\partial \phi_3^2} \right|_v = 8\pi^2 v^2 (B+C) = M_H^2 \; .
\label{``(9)''}
\end{equation}
Finally, the potential (6) is related to its corresponding tree-potential (1) by the requirement \cite{1}
\begin{equation}
\left. \frac{\partial^4 V_{RSB}}{\partial \phi_3^4}\right|_v = \frac{\partial^4 \left[ \lambda_{RSB}(\phi^+ \phi)^2 \right]}{\partial \phi_3^4} = 6\lambda_{RSB}
\label{``(10)''}
\end{equation}
in which case \cite{3}
\begin{equation}
\lambda_{RSB} / 4\pi^2 \equiv y_{RSB} = \frac{11}{3} B + \frac{35}{3} C + 20 D + 16 E \; .
\label{``(11)''}
\end{equation}
Note that the requirement (10) cannot be fulfilled at $v = 0$; \cite{1} spontaneous symmetry breaking must occur.  Since $\{B, C, D, E\}$ are functions of known couplings $g_t(v)$, $g_3(v)$ $g_2(v)$, $g^\prime (v)$ as well as the unknown coupling $\lambda_{RSB}(v)$, eq. (11) leads to a polynomial determination of $\lambda_{RSB}$. \cite{3}

One may utilize eq. (6) to obtain quartic couplings in the Higgs-Goldstone sector for RSB in much the same way that one would utilize the potential of eq. (1) for lowest-order CSB results.  For example, the CSB tree-amplitudes for $H \rightarrow (Z_l Z_L, \; W_L^+ W_L^-)$ assuming $M_H >> M_Z$, respectively are
\begin{equation}
\left. \frac{\partial^3 V_{CSB}}{\partial \phi_3 \partial \phi_4 \partial \phi_4 }\right|_v = 2\lambda_{CSB} v = M_H^2 / v
\label{``(12)''}
\end{equation}

\begin{equation}
\left. \frac{\partial^3 V_{CSB}}{\partial \phi_3 \partial \phi_1 \partial \phi_1} \right|_v + \left.
\frac{\partial^3 V_{CSB}}{\partial \phi_3 \partial \phi_2 \partial \phi_2}  \right|_v = 4 \lambda_{CSB} v = 2 M_H^2 / v \; .
\label{``(13)''}
\end{equation}
Startlingly, the {\it same} results are obtained from the RSB potential (6) via eq. (9) and the eq. (8) minimization requirement that $A = -B/2$:
\begin{equation}
\left. \frac{\partial^3 V_{RSB}}{\partial \phi_3 \partial\phi_4 \partial \phi_4}\right|_v = 4\pi^2 v (2A+3B+2C) = 8\pi^2 v (B+C) = \frac{M_H^2}{v}
\label{``(14)''}
\end{equation}
\begin{equation}
\left. \frac{\partial^3 V_{RSB}}{\partial \phi_3 \partial\phi_1\partial \phi_1}\right|_v +
\left. \frac{\partial^3 V_{RSB}}{\partial \phi_3 \partial \phi_2 \partial \phi_2}\right|_v
= 16\pi^2 v (B+C) = \frac{2M_H^2}{v}
\label{``(15)''}
\end{equation}

A similar equivalence is obtained for $W_L^+ W_L^-$ scattering amplitudes extracted from the Higgs-Goldstone sector.  For conventional electroweak symmetry breaking, the tree-amplitudes for $W_L^+ W_L^- \rightarrow Z_L Z_L$ and $W_L^+ W_L^- \rightarrow W_L^+ W_L^-$ are respectively
\begin{equation}
\frac{\partial^4 V_{CSB}}{\partial w^+ \partial w^- \partial \phi_4 \partial\phi_4} = 2\lambda_{CSB} = M_H^2 / v^2,
\label{``(16)''}
\end{equation}
\begin{equation}
\frac{\partial^4 V_{CSB}}{\partial w^+ \partial w^- \partial w^+ \partial w^-} = 4\lambda_{CSB} = 2M_H^2 / v^2,
\label{``(17)''}
\end{equation}
where $w^\pm \equiv (\phi_1 \pm i \phi_2) / \sqrt{2}$.  The analogous amplitudes, as computed from the potential (6) for radiative electroweak symmetry breaking given the normalization defined by the left-hand sides of eqs. (16,17), are found via eq. (9) to be the same as in CSB:
\begin{eqnarray}
\left. \frac{\partial^4 V_{RSB}}{\partial w^+ \partial w^- \partial \phi_4 \partial \phi_4}\right|_v & = & 8\pi^2 \left( A + \frac{3B}{2} + C\right)\nonumber\\
& = & 8\pi^2 (B+C) = M_H^2 / v^2,
\label{``(18)''}
\end{eqnarray}
\begin{eqnarray}
\left. \frac{\partial^4 V_{RSB}}{\partial w^+ \partial w^- \partial w^+ \partial w^-} \right|_v & = & 8\pi^2 (2A + 3B + 2C)\nonumber\\
& = & 16\pi^2 (B+C) = 2M_H^2 / v^2 \; .
\label{``(19)''}
\end{eqnarray}
These counterintuitive results occur despite the inequality $\lambda_{RSB} >> \lambda_{CSB}$ for the same value of $M_H$.
Moreover, these results are seen to occur independent of the quartic-coupling condition (10);  only the first and second derivatives of the RSB effective potential enter in eqs. (14, 15, 18, 19).  Of course, the relative magnitudes of $\lambda$ {\it are} reflected by $HH \rightarrow HH$ scattering, which for the CSB and RSB cases are respectively
\begin{equation}
\frac{\partial^4 V_{CSB}}{(\partial \phi_3)^4} = 6\lambda_{CSB}
\label{``(20)''}
\end{equation}
\begin{equation}
\left. \frac{\partial^4 V_{RSB}}{(\partial \phi_3)^4}\right|_v = 6\lambda_{RSB} \cong 30 \lambda_{CSB} \; .
\label{``(21)''}
\end{equation}
However, the only other signature scattering amplitudes extracted from eq. (6) which differ from the lowest-order CSB tree amplitudes are those for $(W_L^+ W_L^- , \; Z_L Z_L) \rightarrow HH$.  For conventional electroweak symmetry breaking, these lowest order amplitudes in the Higgs-Goldstone sector are
\begin{equation}
\frac{\partial^4 V_{CSB}}{\partial w^+ \partial w^- \partial \phi_3 \partial \phi_3} = \frac{\partial^4 V_{CSB}}{\partial \phi_4 \partial \phi_4 \partial \phi_3 \partial \phi_3}
=  2\lambda_{CSB} = \frac{M_H^2}{v^2}.
\label{``(22)''}
\end{equation}
The corresponding amplitudes for radiatively broken electroweak symmetry are:
\begin{eqnarray}
\left. \frac{\partial^4 V_{RSB}}{\partial w^+ \partial w^- \partial \phi_3 \partial \phi_3} \right|_v
 =  \left. \frac{\partial^4 V_{RSB}}{\partial \phi_4 \partial \phi_4 \partial \phi_3 \partial \phi_3}\right|_v &=& \pi^2 (8A + 28B + 56C + 48D)\nonumber\\
& = & \pi^2 (24B + 56C + 48D)\nonumber\\
& = & 24\pi^2 (B+C) + \pi^2 (32C + 48D)\nonumber\\
& = & 3M_H^2 / v^2 + \pi^2 (32C + 48D) \; .
\label{``(23)''}
\end{eqnarray}

The result (23) is more than a factor of three larger than the result (22), suggesting scattering rates an order of magnitude larger in an RSB scenario than in a CSB scenario for equivalent Higgs boson masses. This is easily seen if we ignore all Standard Model couplants $x_t (v) \equiv g_t^2 (v) / 4\pi^2 = 0.0253, \; z(v) = (\alpha_s(v)/\pi) = 0.0329, \; r(v) \equiv g_2^2 (v) / 4\pi^2 = 0.0109, \; s(v) \equiv g^{\prime^2} (v) / 4\pi^2 = 0.00324$ except for the dominant scalar field self-interaction couplant $y = \lambda_{RSB} / 4\pi^2$, and then consider the scalar-field theory projection (SFTP) of the effective potential for radiative symmetry breaking taken to the summation of all leading logarithms: \cite{7}
\begin{equation}
V_{SFTP}^{LL} = \pi^2 \phi^4  \left[ y / (1-3y \L) - 0.058531\right] \; .
\end{equation}
One finds for this potential that $y = 0.0541$ and that $M_H = 221$ GeV, in close agreement with leading-logarithm results ($y = 0.0538, \; M_H = 218$ GeV) that are inclusive of the other Standard Model couplants.  One easily finds for this potential that the series in eq. (6) is given by $B = 3y^2$, $C = 9y^3$, $D = 27y^4$, $E = 81y^5$, $\ldots$, in which case the numerical value of the final line of eq. (23) is 2.98, to be compared with the numerical value 0.807 for eq. (22).  Thus the scattering amplitude associated with $W_L^+ W_L^- \rightarrow HH$ via the Goldstone Equivalence theorem is, in this approximation, enhanced in the radiative scenario by a factor of approximately 3.7 relative to the tree amplitude in conventional electroweak symmetry breaking for a $220$ GeV Higgs boson.

The full leading-logarithm expression inclusive of Standard Model coupling constants $g_t(v)$, $g_3(v)$, $g_2(v)$, $g^\prime (v)$ yields a very similar result, demonstrating the usefulness of the above SFTP approximation.  One finds that $M_H = 218$ GeV, $C = 0.00152$, $D = 0.000256$, in which case the numerical value for the full leading logarithm case of the final line of eq. (23) is $2.96$.  This is to be compared with $0.785$ for the final line of eq.(22) for the same $M_H$, yielding a factor of $3.8$ enhancement of the lowest-order-CSB amplitude, and providing a clear signature for radiative electroweak symmetry breaking.

\section*{Acknowledgments}

We are grateful for discussions with V. A. Miransky, R. B. Mann, and D. G. C. McKeon, as well as for financial support from the Natural Sciences and Engineering Research Council of Canada.

\end{document}